\documentclass[aps,prb,twocolumn,showpacs]{revtex4}
\usepackage{graphicx}
\usepackage{amsmath}
\usepackage{amssymb}
\usepackage{times}

\graphicspath{{.}}

\DeclareMathOperator{\acos}{acos}

\DeclareMathOperator{\trace}{Tr}

\begin{document}

\title{Chebyshev Expansion Approach to the AC Conductivity 
  of the Anderson Model}

\author{Alexander Wei{\ss}e}
\affiliation{School of Physics, The University of New South Wales, 
  Sydney NSW 2052, Australia}

\date{August 12, 2004 (revised version)}
  
\begin{abstract}
  We propose an advanced Chebyshev expansion method for the
  numerical calculation of linear response functions at finite
  temperature. Its high stability and the small required resources
  allow for a comprehensive study of the optical conductivity
  $\sigma(\omega)$ of non-interacting electrons in a random potential
  (Anderson model) on large three-dimensional clusters. For low
  frequency the data follows the analytically expected power-law
  behaviour with an exponent that depends on disorder and has its
  minimum near the metal-insulator transition, where also the
  extrapolated DC conductivity continuously goes to zero.  In view of
  the general applicability of the Chebyshev approach we briefly
  discuss its formulation for interacting quantum systems.
\end{abstract}


\maketitle

The numerical calculation of linear response functions is one of the
standard tasks in condensed matter theory and many other areas of
physics. In practice, however, the number of degrees of freedom
usually becomes enormously large and can easily reach $N\approx 10^6$
or more, e.g., for a quantum many body problem. A complete
diagonalisation of such systems and a naive evaluation of linear
response functions is prohibitive is such situations, since the
required time would scale at least as $N^3$. The use and development
of new numerical methods which are {\em linear} in the system size has
therefore become an essential part of current research. In the present
work we follow this line and propose an advanced Chebyshev expansion
method for the calculation of dynamical correlation functions at
finite temperature. It exceeds previous attempts, in particular, since
it requires only a single simulation run for all temperatures and, if
applied to non-interacting fermions, for all chemical potentials.

As a particularly interesting application, we study the optical (AC)
conductivity $\sigma(\omega)$ of non-interacting electrons in a random
potential, which has so far resisted a thorough numerical treatment. The
basic model to describe this kind of problem and many of its features
was proposed by Anderson almost fifty years ago~\cite{An58}, and since
then attracted a considerable amount of analytical, numerical, and
experimental work~\cite{Th74:LR85:KM93b}.  Starting from spinless
fermions $c_i^{(\dagger)}$ which are allowed to hop between
neighbouring sites of a crystal, 
\begin{equation}\label{defham}
  H = -t \sum_{\langle ij\rangle}
  \big( c_i^{\dagger} c_j^{} + c_j^{\dagger} c_i^{} \big)
  + \sum_i \epsilon_i^{} c_i^{\dagger} c_i^{}\,,
\end{equation}
disorder can be introduced in the form of a random, uniformly
distributed local potential $\epsilon_i^{} \in [-W/2, W/2]$
parameterised by the disorder strength~$W$.  Given this Hamiltonian
the question arises, whether its one-particle eigenfunctions span the
entire lattice, thus resembling the Bloch waves known from an ordered
crystal ($W=0$), or are localised in the vicinity of certain lattice
sites. Naturally, this change in the spatial structure of the wave
functions is reflected in the (DC) conductivity of the system, being
insulating or metallic depending on the disorder strength $W$, the
spatial dimension $d$, and the particle density (or chemical potential
$\mu$). Much of our current understanding of this disorder-induced
metal-insulator transition is based on the one-parameter scaling
theory of Abrahams et~al.~\cite{AALR79}, which in $d\le 2$ dimensions
predicts insulating behaviour for any finite disorder $W>0$ and a
continuous metal-insulator transition at some $W_c(\mu)>0$ for $d>2$.
The critical behaviour near the transition is usually described in
terms of nonlinear $\sigma$-models~\cite{Ef83} and is widely believed
to follow power laws with a correlation/localisation length $\xi$
diverging as $\xi \propto |W_c-W|^{-\nu}$, and the DC conductivity
vanishing as $\sigma(0) \propto (W_c-W)^{s}$. Numerical work confirmed
much of this general picture and over the last years focused on the
precise determination of the critical line $W_c(\mu)$ and of the
critical exponents, which so far could not be calculated analytically.
For the above model the most reliable data ($W_c(0)/t = 16.54$ and
$\nu=1.57$, cf.  Ref.~\onlinecite{SO99}) is based on the transfer-matrix
method~\cite{MK83}, where in a quasi-one-dimensional geometry the
correlation length $\xi$ is obtained from the finite size scaling of
the Lyapunov exponents. Unfortunately, approaches of this type cannot
directly access the DC conductivity $\sigma(0)$ or its critical
behaviour. Our knowledge of the exponent $s$ is therefore mainly based
on scaling arguments~\cite{We76}, namely, $s=(d-2) \nu$.  However, the
validity of the one-parameter scaling theory and of the corresponding
critical behaviour has been repeatedly called into
question~\cite{EV92,Zi86}, and instead the non-power-like critical
behaviour known for the Bethe lattice has been proposed to hold also
for hyper-cubic systems. The resolution of this certainly not
completely settled issue may require the use of alternative numerical
methods, which should preferably be based on true $d$ dimensional
systems and yield complementary critical quantities.

As noted before, here we want to focus on the numerical calculation of
the optical conductivity $\sigma(\omega)$ of three-dimensional (cubic)
clusters. This allows for a test of various analytical predictions for
the finite frequency behaviour, and eventually we can draw
conclusions about the zero-frequency response. In particular, for $d$
dimensional systems Wegner~\cite{We76} found $\sigma(\omega) \sim
\omega^{(d-2)/d}$ to hold exactly at the metal-insulator transition, a
prediction which is consistent also with the one-parameter scaling
theory~\cite{SA81}. On the metallic side of the transition different
studies~\cite{SA81,OW79,Sh82} agree that for small enough frequency
the conductivity should behave as $\Delta\sigma = \sigma(\omega) -
\sigma(0) \sim \omega^{(d-2)/2}$, whereas on the insulating side we
expect the well known $\sigma(\omega)\sim \omega^2$ behaviour
independent of the spatial dimension~\cite{Mo67}.  As will become
clear below, the numerical calculation of $\sigma(\omega)$ is a
challenging task, which certainly is the reason that only the
prediction for the critical point in $d=3$, i.e., $\sigma(\omega)\sim
\omega^{1/3}$, is confirmed so far~\cite{LS94b,SN99}.  Within linear
response the real part of the optical conductivity is given by
\begin{equation}\label{defsigma}
  \sigma(\omega) = \sum_{n,m}
  \frac{|\langle n|J_x| m\rangle|^2}{\omega L^d} [f(E_m) - f(E_n)]\,
  \delta(\omega - \omega_{nm})\,,
\end{equation}
where $|n\rangle$ and $|m\rangle$ denote eigenstates of the
Hamiltonian with energies $E_n$ and $E_m$, $\omega_{nm}=E_n-E_m$,
$f(E) = 1/(\exp(\beta(E-\mu)) + 1)$ is the Fermi function, and $J_x =
-\text{i} t \sum_{i} (c_i^{\dagger} c_{i+x}^{} - c_{i+x}^{\dagger}
c_i^{})$ the $x$-component of the current operator. Even at zero
temperature Eq.~\eqref{defsigma} involves a summation over matrix
elements between {\em all} one-particle eigenstates of $H$, which can
hardly be calculated for a reasonably large system. Consequently,
until now, the number of numerical attempts to this problem is very
small. Some authors relied on a full diagonalisation of the
Hamiltonian and an explicit summation of the current matrix
elements~\cite{LS94b,AG78,SM85,HGF87}, but of course the system sizes
manageable with this approach are very limited. Even the dramatically
improved performance of present day computers allows only the study of
clusters of about $L^3=20^3$ sites. More recently the so-called forced
oscillator method~\cite{SN99} and the projection method~\cite{Ii98}
were applied to the problem, which increased the accessible system
size to about $30^3$ and $256^3$ sites, respectively. However, the
frequency and parameter ranges considered in these works were rather
limited, and unfortunately the resolution as well as the statistical
quality of the data seem to be insufficient for a detailed analysis of
the low-frequency behaviour~\cite{Ii98}.

About a decade ago Silver and R\"oder~\cite{SR94} proposed the kernel
polynomial method (KPM) for the calculation of the density of states
of large Hamiltonian matrices, which, in addition, turned out to be a
very robust and reliable tool for the calculation of temperature
dependent static quantities and zero-temperature dynamical correlation
functions of interacting systems (which in contrast to
Eq.~\eqref{defsigma} require only a single summation over the matrix
elements between the ground-state and excitations)~\cite{SRVK96}. In a
nutshell, after appropriate rescaling of the Hamiltonian, $\tilde H =
(H-b)/a$, and of the energy spectral quantities like the density of
states, $\rho(E) = \sum_{n=0}^{N-1} \delta(E - E_n)/N$, are expanded
in terms of Chebyshev polynomials $T_m(x) = \cos(m \acos(x))$. To
alleviate the effects of a truncation of such a series the result is
convoluted with a particular kernel (the Jackson kernel), and to a
good approximation $\rho(E)$ then reads
\begin{equation}
  \rho(E) \approx 
  \frac{g_0 \mu_0 + 2 \sum_{m=1}^{M-1} g_m \mu_m \ T_m[(E-b)/a]
  }{\pi \sqrt{a^2 - (E-b)^2}}
\end{equation}
Here the $g_m$ account for the kernel and the $\mu_m$ are the actual
expansion coefficients, $\mu_m = \int \rho(x)\, T_m[(x-b)/a]\, dx =
\trace[ T_m(\tilde H)]/N$.  It turns out that the numerical
calculation of the coefficients $\mu_m$ does not require the full
evaluation of the trace of the polynomial $T_m(\tilde H)$. Instead,
self-averaging properties, used also in Monte Carlo simulations, allow
for an replacement of the trace by an average over a small number
$R\ll N$ of random states $|r\rangle$. If, in addition, recursion
relations for the Chebyshev polynomials are taken into account, for
sparse Hamiltonians of dimension $N$ the numerical effort for the
calculation of all $M$ coefficients $\mu_m$ is proportional to
$RNM/2$, i.e., {\em linear} in $N$. Once the $\mu_m$ are known the
reconstruction of the target function is facilitated by the close
relation between Chebyshev expansion and Fourier transform, i.e., the
availability of divide-and-conquer type algorithms (FFT).

\begin{figure}
  \includegraphics[width=0.9\linewidth]{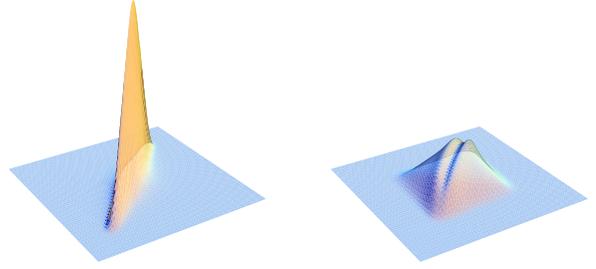}
  \caption{The matrix element density $j(x,y)$ for 
    the Anderson model at $W/t=2$ and $12$. Note the dip developing at
    $x=y$ which finally causes the vanishing DC
    conductivity.}\label{figjxy}
\end{figure}

So far we are aware of only one attempt~\cite{IE03} to generalise the
kernel polynomial method to finite-temperature dynamical correlations
(note that for non-interacting systems the numerical effort is equal
for $T=0$ and $T>0$). In this recent letter Iitaka and
Ebisuzaki~\cite{IE03} propose a Chebyshev expansion of the Boltzmann
or Fermi weights (see Eq.~\eqref{defsigma}), which is used to generate
a set of correspondingly weighted random vectors. These states are
then subject to standard numerical time evolution and measurements of
the targeted operator, and finally yield the considered correlation
function. Although certainly being a useful approach, we argue that it
is still unnecessarily complicated, mainly because each change in the
temperature $T$ or chemical potential $\mu$ requires a new simulation.

To avoid these complications we propose a slight increase in the level
of abstraction, namely, the introduction of {\em two-dimensional} KPM.
A closer inspection of Eq.~\eqref{defsigma} shows that
$\sigma(\omega)$ is easily written as an integral over a matrix
element density
\begin{equation}\label{defjxy}
  \begin{aligned}
    j(x,y) & = \frac{1}{L^d} \sum_{n,m} |\langle n|J_x| m\rangle|^2 
    \delta(x - E_n) \delta(y - E_m)\,,\\
    \sigma(\omega) & = \frac{1}{\omega} 
    \int\limits_{-\infty}^{\infty} j(x, x+\omega) [f(x) - f(x+\omega)]\, dx\,.
  \end{aligned}
\end{equation}
The quantity $j(x,y)$, however, is of the same structure as the
density of states, except for being a function of two variables.  As
was shown by Wang~\cite{WZ94:Wa94} some years ago, it can thus be
expanded as a series of polynomials $T_l(x) T_m(y)$ and the expansion
coefficients $\mu_{lm}$ are characterised by a similar trace,
$\mu_{lm} = \trace[T_l(\tilde H) J_x T_m(\tilde H) J_x]/L^d$.  Again
the trace can be replaced by an average over just a few random vectors
$|r\rangle$, and the numerical effort for an expansion of order
$l,m<M\ll N$ ranges between $2RNM$ and $RNM^2$, depending on whether
memory is available for up to $M$ vectors of dimension $N$ or not.
Probably overlooking the potential of the approach, so far only the
zero temperature response was studied and, in particular, the back
transformation of the expansion coefficients relied on pure truncated
Chebyshev series~\cite{WZ94:Wa94}. The latter, however, suffer from
unwanted high-frequency oscillations and the positivity of $j(x,y)$ is
not ensured. We therefore generalised the Jackson kernel and the KPM
to two dimensions. Combined with fast Fourier methods, which are
available for arbitrary dimension, this leads to an easy and reliable
method for the calculation of $j(x,y)$ and $\sigma(\omega)$.

Note the main advantage of this approach: Once we know the
coefficients $\mu_{lm}$ and the resulting $j(x,y)$, we can immediately
calculate $\sigma(\omega)$ for {\em all} temperatures and {\em all}
chemical potentials, without repeating the most time consuming step of
calculating $\mu_{lm}$ (and, for the present model, averaging over
several realisations of disorder). In addition, as was shown in a
number of works, standard KPM is numerically much more stable and
allows much higher resolution than the popular Lanczos recursion
approach~\cite{HHK72}. We therefore believe that the new
generalisation of KPM will also outperform the finite-temperature
Lanczos methods proposed recently~\cite{JP94,ADEL03}. The
generalisation of the approach to interacting systems is
straightforward~\cite{footnote}. It merely requires a substitution of
the Fermi function by the Boltzmann weight in Eq.~\eqref{defjxy}, and
a division of the result by the partition function, which is readily
obtained from an expansion of the density of states.

\begin{figure}
  \includegraphics[width=\linewidth]{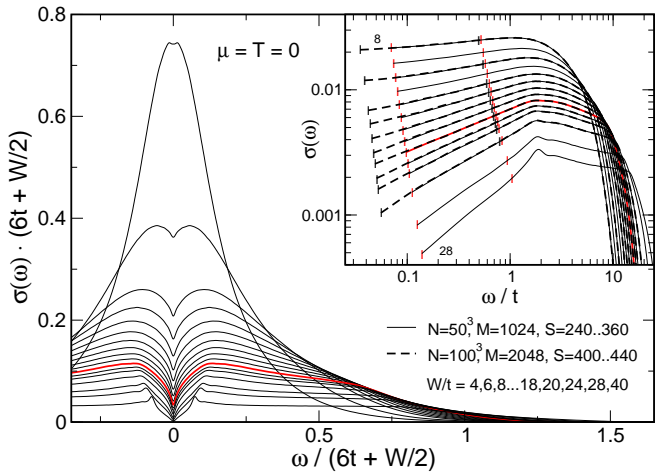}
  \caption{Optical conductivity of the 3D Anderson 
    model at $T=0$ and $\mu = 0$ (band centre) for increasing disorder
    $W$. The thick red lines mark $W/t = 16$, which approximately
    corresponds to the critical disorder. Data denoted by solid lines is
    based on $N=50^3$ site clusters, expansion order $M=1024$, and
    $S=240\ldots 360$ disordered samples, dashed lines in the inset
    correspond to $N=100^3$, $M=2048$ and $S=400\ldots 440$.}\label{figover}
\end{figure}

Applying the approach to the Anderson model, we obtain the matrix
element density $j(x,y)$ shown in Figure~\ref{figjxy}. Starting from a
``shark fin'' at weak disorder, with increasing $W$ the density
$j(x,y)$ spreads in the entire energy plane, simultaneously developing
a sharp dip along $x=y$. A comparison with Eq.~\eqref{defjxy} reveals,
that it is this dip which is responsible for the decreasing and
finally vanishing DC conductivity. For $\mu=0$ (band centre) and $T=0$
the corresponding optical conductivity $\sigma(\omega)$ is given in
Figure~\ref{figover}. Note, that the calculation is based on large
finite clusters with up to $N = L^3 = 100^3$ sites and periodic
boundary conditions, the data is averaged over up to $S=440$
disordered samples, and the expansion order $M=1024$ (or $M=2048$ for
the dashed sets in the inset).  At weak disorder the conductivity is
almost Drude like with only a small dip at low frequency. With
increasing disorder this small-$\omega$ feature becomes more
pronounced and finally leads to insulating behaviour at strong
disorder. Beyond a sharpening maximum near $\omega\approx t$ the
conductivity falls of almost with a power law and later exponentially.

\begin{figure}
  \begin{center}
    \includegraphics[width=\linewidth]{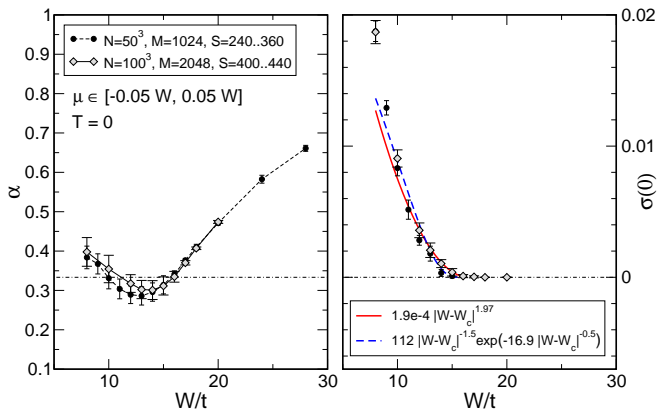}
  \end{center}
  \caption{Exponent $\alpha$ and DC conductivity 
    $\sigma(0)$ obtained from fits of the low-frequency conductivity
    to $\sigma(\omega) = \sigma(0) + C\,\omega^{\alpha}$ (vertical
    bars in the inset of Fig.~\ref{figover} mark the underlying
    frequency range). Error bars are estimated by slightly varying
    $\mu$ in the range $-0.05 W\ldots 0.05 W$.  }\label{figfits}
\end{figure}

The high precision of the data allows for a detailed comparison of the
low frequency behaviour with the above mentioned analytical results.
In the inset of Figure~\ref{figover} we focus on the low frequency
part and plot the conductivity data again on a double-logarithmic
scale. Clearly, for disorder $W/t\ge 16$ the data follows a power law,
whereas for $W/t<16$ the slight upturn at low frequencies accounts for
the finite DC conductivity. To substantiate these findings, in
Figure~\ref{figfits} we show fits of the low-frequency data to
$\sigma(\omega) = \sigma(0) + C \omega^{\alpha}$. Starting from the
localised phase at large $W$ the DC conductivity $\sigma(0)$ is zero
and the exponent $\alpha$ decreases continuously with $W$, reaching
$\alpha = 1/3$ near $W/t\approx 16$. Below that value $\sigma(0)$
increases continuously with decreasing disorder $W$, and the same
seems to hold for $\alpha$. Note that we slightly vary $\mu$ around
zero to expand the data basis and estimate the error of the fits.
Unfortunately, for $W/t<16$ the three free parameters lead to a
sizeable uncertainty in particular for the exponent $\alpha$.
Nevertheless, we can confirm the general trends, namely an increase of
the exponent $\alpha$ from $1/3$ at the critical point to eventually a
value of $2$ at very large disorder, and an increase towards $\alpha =
(d-2)/2 = 1/2$ for weak disorder.  Although our data looks rather
convincing, note one potential problem: The considered frequencies
might still be too large for an observation of the correct scaling,
since from analytical work~\cite{Sh82} the $\sqrt{\omega}$ or
$\omega^2$ behaviour of $\sigma(\omega)$ is expected only for
frequencies smaller than a cut-off of the order of $\omega_{\text{cr}}
\sim 1/(\rho(\mu) \xi^3)$, while for $\omega\gg \omega_{\text{cr}}$
$\Delta\sigma \sim \omega^{1/3}$. On the other hand, also an increased
resolution did not show any indication of such a cross-over, even
though, particularly on the insulating side, the localisation length
$\xi$ rapidly decreases with $W$, reaching the order of $1$ for the
largest disorder values considered. We hope further studies can
resolve this puzzling issue.

Keeping in mind the above subtleties, we can also try to address the
critical behaviour expressed in $\sigma(0)$. As the comparison of data
for $50^3$ and $100^3$ sites in Figure~\ref{figover} illustrates, for
the considered frequencies the AC conductivity does not suffer from
noticeable finite-size effects. This is corroborated by estimates of
the diffusion length $L_{\omega}$ (the distance electrons diffuse
within a field cycle; cf.  Ref.~\onlinecite{SA81}), throughout yielding
$L_{\omega}\ll L$.  Therefore the fit parameter $\sigma(0)$ in
Figure~\ref{figfits} should correspond to the thermodynamic limit of
the DC conductivity, which for dimension $d=3$ is widely believed to
follow a $\sigma(0) \sim (W_c - W)^{s}$ law with $s = \nu\approx
1.57$.  However, the curvature of $\sigma(0)$, derived from our data,
seems to be larger, leading to $s$ of the order of~$2$.  On the other
hand, we also obtained reasonable fits using the expression for the
Bethe lattice~\cite{EV92}, $\sigma(0) \sim (W_c - W)^{-3/2} \exp(-A
(W_c - W)^{-1/2})$, which would contradict the behaviour generally
assumed for the $d=3$ Anderson model.  Although resolving these
interesting questions certainly requires an improvement of both the
resolution and the statistical quality of the data, our results shed
new light on the Anderson transition and illustrate the potential of
the numerical approach.

In summary, we described a promising new technique for the numerical
calculation of finite temperature dynamical correlation functions for
both interacting and non-interacting quantum systems. By extending the
Kernel Polynomial Method to functions of two variables, we avoid the
disadvantages of thermal projection techniques, and obtain reliable
results for all temperatures (and chemical potentials) from a single
simulation run. Being a hybrid of the iterative schemes of numerical
diagonalisation and of random sampling, the approach might also
inspire new Monte-Carlo methods for correlation functions. Applying
the method to the Anderson model we present comprehensive data for the
AC conductivity, which substantially improves previous numerical
studies with respect to accessible system size, considered frequency
and parameter range, as well as statistical significance. In addition,
we confirm analytical predictions for the low-frequency behaviour of
the AC conductivity, but find indications that the critical behaviour
of the DC conductivity might deviate from the commonly presumed form.

{\small The author acknowledges valuable discussion and comments of
  H.~Fehske and J.~Oitmaa, the hospitality at the MPI PKS Dresden, the
  grant of computational resources by APAC and ac3, and financial
  support by the Australian Research Council.}


\end{document}